\documentclass[conference]{IEEEtran}
\usepackage{graphicx}
\usepackage{amssymb}
\usepackage{epstopdf}

\usepackage[ruled,vlined]{algorithm2e}
\usepackage{amsmath, amsthm, amssymb}
\usepackage{shadethm}
\numberwithin{equation}{section}
\usepackage{rotating}
\usepackage{subfigure}
\usepackage{hyperref}
\usepackage{float} 
\usepackage{soul} 
\usepackage{todonotes}
\usepackage{wrapfig}
\usepackage{rotating}
\usepackage{xcolor}
\usepackage{fancyhdr}
\usepackage{array}
\usepackage{numprint}

\graphicspath{{graphics/}}

\newcommand{\algo}[1]{\textsf{#1}}
\newcommand{\term}[1]{\emph{#1}}
\newcommand{\atitle}[1]{\textsl{#1}}
\newcommand{\func}[1]{ \mathit{#1} }

\newcommand{\NP}{$\mathcal{N}\mathcal{P}$}
\newcommand{\auth}{\smile}
\newcommand{\coauth}{\frown}
\newcommand{\mb}[1]{\mathbf{#1}}

\newcommand{\AND}{\wedge}

\newcommand{\pairs}[1]{\textstyle{#1 \choose 2}}
\newcommand{\set}[1]{ \{ #1 \} }

\newcommand{\ie}{i.e.\ }
\newcommand{\eg}{e.g.\ }
\newcommand{\etal}{et\! al.\! }

\newcommand{\margindef}[1]{\marginpar{\textsf{{\scriptsize #1}}} } 

\makeatletter
\@addtoreset{paragraph}{section}
\makeatother

\theoremstyle{slplain}

\newtheorem{definition}{Def.}

\newcount\shortyear\newcount\shorthour\newcount\shortminute
\shorthour=\time\divide\shorthour by 60\shortyear=\shorthour
\multiply\shortyear by 60\shortminute=\time\advance\shortminute by
-\shortyear
\shortyear=\year\advance\shortyear by -1900

\def\zeit{\number\shorthour:\ifnum\shortminute<10 0\number\shortminute
\else\number\shortminute\fi}

\begin{document}
%
\title{Static and Dynamic Aspects of \\ Scientific Collaboration Networks}

\author{\IEEEauthorblockN{Christian Staudt}
\IEEEauthorblockA{\small{christian.staudt@student.kit.edu}}
\and
\IEEEauthorblockN{Andrea Schumm}
\IEEEauthorblockA{\small{andrea.schumm@kit.edu}}
\and
\IEEEauthorblockN{Henning Meyerhenke}
\IEEEauthorblockA{\small{meyerhenke@kit.edu}}
\and
\IEEEauthorblockN{Robert G\"{o}rke}
\IEEEauthorblockA{\small{robert.goerke@kit.edu}}
\and
\IEEEauthorblockN{Dorothea Wagner}
\IEEEauthorblockA{\small{dorothea.wagner@kit.edu}}
\and
\IEEEauthorblockA{
Institute of Theoretical Informatics, Karlsruhe Institute of Technology (KIT), Am Fasanengarten 5, 76131 Karlsruhe, Germany
}
}

\maketitle


\begin{abstract}
Collaboration networks arise when we map the connections between scientists which are formed through joint publications.
These networks thus display the social structure of academia, and also allow conclusions about the structure of scientific knowledge.
Using the computer science publication database \atitle{DBLP}, we compile relations between authors
and publications as graphs and proceed with examining and quantifying collaborative relations with
graph-based methods.
We review standard properties of the network and rank authors and publications by centrality. Additionally, we detect communities with \term{modularity}-based clustering and compare the resulting clusters to a ground-truth based on conferences and thus topical similarity.
In a second part, we are the first to combine \atitle{DBLP} network data with data from the \term{Dagstuhl Seminars}:
We investigate whether seminars of this kind, as social and academic events designed to connect researchers, leave a visible track in the structure of the collaboration network.
Our results suggest that such single events are not influential enough to change the network structure significantly.
However, the network structure seems to influence a participant's decision to accept or decline an invitation.
\end{abstract}

%
\IEEEpeerreviewmaketitle

\section{Introduction}
In scientometrics, the quantitative study of science, network analysis has become a prominent tool. \term{Coauthorship networks} have attracted interest both as \term{social networks} and as \term{knowledge networks}: They display the social structure of academia, while their bibliographic aspect allows conclusions about the structure of scientific knowledge.
Accordingly, networks of this kind are the objects of ongoing research: Newman~(\cite{Newman2001a,Newman2004,Newman2001-SCN-I,Newman2001-SCN-II}), for example, studies properties of coauthorship networks in the realm of physics (\atitle{Los Alamos e-Print Archive}, \atitle{SPIRES}), mathematics (\atitle{Mathematical Reviews}), biomedical research (\atitle{Medline}) and computer science (\atitle{NCSTRL}), summarizing many statistical properties of coauthorship networks. Aspects like \term{connectedness}, \term{distance}, \term{degree distribution}, \term{centrality} and \term{community structure} are recurring themes in such studies. Where we follow up on these topics, we cite relevant related work in the respective sections of this paper.

Based on the extensive publication database \term{DBLP}~\cite{dblp}, we model relations between authors and publications as graphs, mapping almost the entire field of computer science. This allows us to examine and quantify the collaborative relations between researchers using graph-based methods. We compile a graph in which edges link coauthors, as well as a bipartite author-paper graph. In the first part, we review standard properties of the network, rank authors and publications by centrality, and detect communities with \term{modularity}-based clustering. In the second part, we combine the network with seminar data provided by the \term{Schloss Dagstuhl}~\cite{dagstuhl} conference center: The \term{Dagstuhl Seminars} assemble researchers with the goal of fostering (collaborative) work in cutting-edge areas of computer science. We examine whether such events leave a track in the structure of the collaboration network. For this purpose, we apply appropriate measures to a time-resolved version of the \term{authorship graph}.

We are the first to perform a joint analysis of the \atitle{Dagstuhl} and \atitle{DBLP} datasets, which allows us to study the impact of social/academic events on the time-evolution of the network structure.
Our results suggest that a participant's decision to accept or decline an invitation can be predicted from the network data to some extent.
While our analysis of the \atitle{DBLP} data mostly confirms properties of similar networks, the distribution of the number of coauthors differs from data reported in~\cite{Newman2001-SCN-II}.
We also describe an approach to finding central researchers based on \term{eigenvector centrality} in the bipartite authorship graph, a combination that to our knowledge has not been used before.
Additionally, we apply \term{modularity} clustering and compare the detected communities to a ground-truth defined by conferences, from which we infer distinct areas of research.

\section{Preliminaries}

\subsection{Collaboration Network Model}

 As of 2011, \atitle{DBLP} covers about 1.5 million publications by 0.8 million authors. The earliest work dates from 1936, and we include all works up to 2009 in our analysis. We describe briefly how a coauthorship network is extracted from the publication database and represented as different types of graphs. The database associates publications and authors and thus provides two main relations, \term{authorship} and \term{coauthorship}, formalized as follows: 

\begin{definition}
Given the sets of authors $\mb{A}$ and publications $\mb{P}$, the \term{authorship} relation is defined as
\begin{equation*}
	\forall \{a, p \} \in \mb{A} \times \mb{P}: \quad a \auth p \iff \text{$a$ is author of $p$}
\end{equation*}

The \term{coauthorship} relation between two authors from $\mb{A}$ is defined as
\begin{equation*}
	\forall \{a, b \} \in \mb{A} \times \mb{A}: a \coauth b \iff  \exists p \in \mb{P}: a \auth p  \AND  b \auth p
\end{equation*}

\end{definition}

From these, two graph representations of the network follow: A bipartite \term{authorship graph} (or author-paper graph) $G_\mb{PA}$, in which each publication is connected by edges to its authors; and a \term{coauthorship graph} $G_\mb{A}$, in which two authors are connected by an edge if they are coauthors of a joint publication. 
 
 \begin{definition}
The \term{authorship graph} is a mapping from the sets of publications $\mb{P}$ and authors $\mb{A}$ to the node sets $V_{\mb{P}}$ and $V_{\mb{A}}$, resulting in a bipartite graph $G_{\mb{PA}} = (V_{\mb{A}}, V_{\mb{P}}, E)$, where
\begin{equation*}
\{v_a, v_p\} \in E \iff a \auth p
\end{equation*}
\end{definition}%

\begin{definition}
The \term{coauthorship graph} is a mapping from the set of authors $\mb{A}$ to the node set $V_\mb{A}$, resulting in the graph $G_\mb{A} = (V_\mb{A}, E)$, where
\begin{equation*}
	\{v_a, v_b\} \in E \iff a \coauth b
\end{equation*}
\end{definition}%

While $G_\mb{A}$ is sufficient when focusing only on the social network of coauthors, $G_\mb{PA}$ preserves the publications as the cause of relations, as well as single-author publications. Table~\ref{size-graphs} shows the size of the graphs constructed from the full publication data set. 

\begin{table}[htdp]
\begin{center}
\begin{tabular}{|c|c|c|}
\hline
 graph & $n$ & $m$ \\
 \hline
$G_\mb{PA}$ & 2 296 586 & 3 775 881 \\
$G_\mb{A}$ & 852 250 & 2 785 037 \\
\hline
\end{tabular}
\end{center}
\caption{Size of resulting graphs}
\label{size-graphs}
\end{table}%

In order to determine whether events have effects detectable in terms of the network (Section~\ref{sec:impact}), we also track groups of authors over the course of time, using a sequence of graphs in which each graph represents a current snapshot of the authorship relations. This \term{time-resolved} version of $G_\mb{PA}$ enables us to study the dynamics of the network: Let $t(p)$ denote the publication date of publication $p$. Then the publications from a time segment $[y,z], z > y$, are
\begin{equation*}
\mb{P}_{[y,z]} := \set{ p \in \mb{P}: y \le t(p) \le z}
\end{equation*}
 The respective authors of these publications are
 \begin{equation*}
 \mb{A}_{[y,z]} := \set{a \in \mb{A}: \exists p \in \mb{P}_{[y,z]}: a \auth p}
 \end{equation*}
The graph sequence is constructed on the basis of a sliding time segment, with parameters width $w$ and increment $s$:

\begin{definition}
The \term{time-resolved authorship graph} is a sequence of graphs $\mathcal{G}_{\mb{PA}}^{w,s}$ where each graph in the sequence is constructed from the publications in $\mb{P}_{[y,y+w]}$ and the authors up to $\mb{A}_{[y,y+w]}$ using a sliding time segment with width $w$ and increment $s$.
\end{definition}

Author nodes are aggregated over time, while publications are deleted for each step in the sequence.
A time segment and increment of 1 year was chosen for the study in Section~\ref{sec:impact}, the finest time resolution possible with \atitle{DBLP} data.

\section{Network Properties and Community Structure}

\subsection{General Network Properties}

We briefly review some general properties of the collaboration network:

\paragraph{Connectedness}

$G_\mb{A}$ features a \term{giant connected component} containing about 80\% of all authors. (Giant components connecting up to 90\% of all authors have previously been detected across scientific fields~\cite{Newman2001-SCN-I}). Aside from the 6\% of the authors without collaborations, about 14 \% of author nodes are distributed over a multitude of small components with few publications.
We conclude that, in general, authors who have worked on several publications and were part of more than one collaborative team join the large connected component.
In terms of average distances between researchers (6.58 for a sample), we confirm the previously reported \term{small world} property for the field of computer science~\cite{Newman2001-SCN-II}  and \atitle{DBLP} in particular~\cite{Elmacioglu2005}.

\paragraph{$k$-Core Structure} 

\begin{figure}[htbp]
\begin{center}
\includegraphics[width=0.43\textwidth]{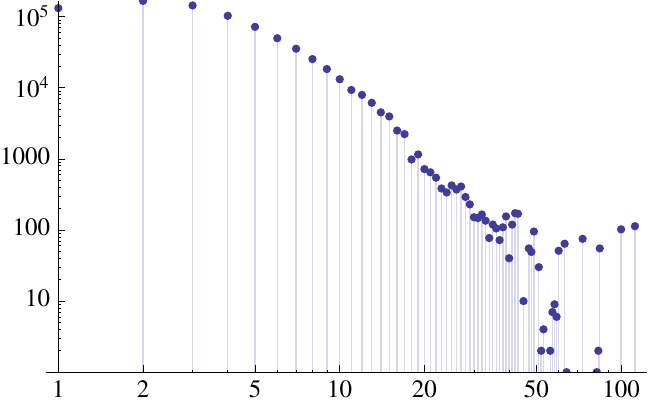}
\caption{Histogram of \term{core numbers} in $G_\mb{A}$ (x-axis: \term{core number}, logarithmic y-axis: frequency)}
\label{fig:core-numbers}
\end{center}
\end{figure}

A $k$-core is a maximal subgraph in which each node is adjacent to at least $k$ other nodes. $k$-cores refine the concept of connected components (which form the 1-core);
\term{$k$-core decomposition} reveals nested, successively more cohesive layers of the graph. We assign each node a \term{core number}, the highest $k$ for which there is a $k$-core containing the node.
Figure~\ref{fig:core-numbers} shows a histogram of the resulting \term{core numbers} in $G_\mb{A}$ with two logarithmic axes. The rather uniform sequence indicates uniform density and cohesiveness of the graph, showing that the network does not have strongly cohesive groups of authors embedded in shells of weakly connected authors~\cite{scott-sna}. 
A more extensive $k$-core analysis of a \atitle{DBLP}-based coauthorship network is presented in~\cite{Giatsidis}.

\paragraph{Degree Distribution}

\begin{figure}[htbp]
\begin{center}
\includegraphics[width=0.85\columnwidth]{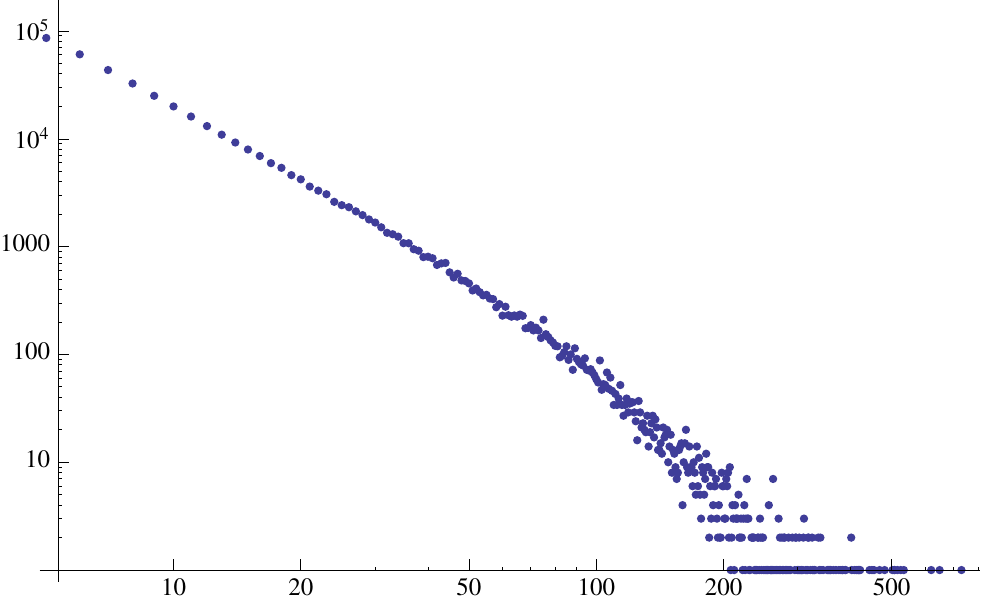}
\caption{Degree distribution in $G_A$ (logarithmic x-axis: degree, logarithmic y-axis: frequency)}
\label{default}
\end{center}
\end{figure}

Node degree in $G_\mb{A}$ corresponds to the number of coauthors of each author.
The degree distribution is highly skewed.
It indicates a \term{scale-free network}, in which the frequency $P(k)$ of nodes with degree $k$ follows a power law, \ie $P(k) \sim k^{-\gamma}$, with coefficient $\gamma = 2.889$.
Newman~\cite{Newman2001-SCN-I} reports a differing power-law degree distribution in the number of coauthors with $\gamma = 3.41$ for computer science, based on \atitle{NCSTRL}.

\paragraph{Summary} General properties indicate that the network of collaborations in computer science is in many respects a typical social network: It shows participation inequality (visible as a power-law degree distribution), with a few highly prolific authors and many smaller contributions. It also features a high degree of connectedness, a giant component, and mostly short paths between arbitrary pairs of nodes. Our observations are in agreement with the results of related studies (except for the degree power-law exponent), indicating that these properties are universal features of scientific collaboration networks.

\subsection{Centrality}

\term{Centrality measures} were formulated to identify nodes which are structurally prominent or influential, due to their position in the center of a network. \term{Betweenness} and \term{closeness centrality} have previously been applied to coauthorship graphs with the goal of identifying influential scientists in their respective fields~(\cite{Newman2001-SCN-II, Boerner2005}).
Elmacioglu \etal report a ranking of prominent scholars by \term{closeness} and \term{betweenness} centrality~\cite{Elmacioglu2005}.
As a rationale, it has been stated that authors with high \term{betweenness} are important intermediates for interactions or information flows, as it allows them to control such flows; high closeness is assumed to be an advantage for accessing or disseminating information~\cite{Elmacioglu2005}. 
However, it is not clear why academic influence should be understood mainly as the ability to mediate interactions. Furthermore, the network of information flow in academia and the network of coauthorship relations may be quite distinct.
We therefore follow a different approach based on \term{eigenvector centrality}~\cite{b-fawassci-72} in the bipartite authorship graph:
It assumes that an author's influence is first of all proportional to the amount of publications. Additionally, the contribution of a paper to an author's centrality should be weighted depending on the centrality of the coauthors.

\begin{definition}
\term{Eigenvector centrality}: Given a graph $G$ with adjacency matrix $A$, we require a centrality score $x_i$ of node $v_i$ to be proportional to the scores of its neighbors:
$$x_i = c \sum_{j=1}^n A(i,j) \ x_j \qquad c \neq 0$$

By the Perron-Frobenius theorem, there exists a nonnegative eigenvector $x$ of $A$ (satisfying $A x = \frac{1}{c} x = \lambda x$) which corresponds to the largest eigenvalue $\lambda$. An entry $x_i$ constitutes the desired centrality score for vertex $v_i$.
\end{definition}

Modeling the collaboration network as the bipartite graph $G_\mb{PA}$ has the benefit that it allows us to assign a centrality score to a publication as a node, rather than just account for a publication as an edge attribute or weight in $G_\mb{A}$~\cite{bonacich2004hyper}.
Thus, our centrality scores express the concept that authors are central in the collaboration network to the extent that they have collaborated on central publications with other central authors.
In this respect, the approach is similar to ranking webpages with the \algo{PageRank} algorithm, where hyperlinks are treated as votes to the relevance of the target page and are weighted by the relevance of the source page.

\begin{table}
\vspace{3ex}
\begin{small}
$$
\begin{array}{|c|c|}
\hline
\text{centrality} \cdot 10^{-5} & \text{author} \\
\hline
9.76232 & \text{Diane Crawford} \\
9.45441 & \text{Robert L. Glass} \\
9.08697 & \text{Chin-Chen Chang} \\
8.30777 & \text{Edwin R. Hancock} \\
7.91401 & \text{Grzegorz Rozenberg} \\
7.82901 & \text{Joseph Y. Halpern} \\
7.75409 & \text{Sudhakar M. Reddy} \\
7.69387 & \text{Philip S. Yu} \\
7.50894 & \text{Moshe Y. Vardi} \\
7.47370 & \text{Ronald R. Yager} \\
 \hline
 \end{array}
 $$
 \end{small}
 \caption{Top segment of author ranking by centrality}
  \label{tab:ranking-authors}
  \vspace{1ex}
 \end{table}
 %
 
%
 
\begin{figure}[bt]
\begin{center}
\subfigure[authors]{
      \includegraphics[width=.35\textwidth]{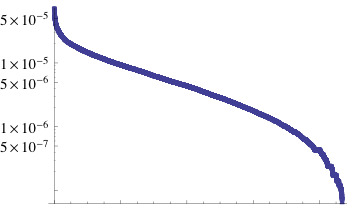}
      \label{fig:centrality-authors}
    }
    \subfigure[publications]{
      \includegraphics[width=.35\textwidth]{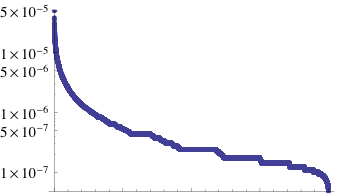}
      \label{fig:centrality-publications}
    }
  \caption{Centrality scores (logarithmic y-axis), sorted}
  \label{fig:centrality-authors-publications}
  \end{center}
\end{figure}

Figure~\ref{fig:centrality-authors-publications} shows the distributions of centrality scores for authors and publications. Extreme values are less frequent, and the distribution does not exhibit a power law.
Table~\ref{tab:ranking-authors} contains the top segment of an author ranking by our approach to centrality.
(See~\cite{ProlificDBLPAuthors} for a comparison to a purely productivity-based ranking of \atitle{DBLP} authors.)
The respective ranking of publications places papers with unusually high author counts at the top, \eg work on large supercomputing and database projects, and further study would be needed to interpret publication centrality properly.
With respect to the evaluation in Section~\ref{sec:impact}, it should also be noted that \atitle{Dagstuhl} seminar invitees have a significantly higher median \term{eigenvector centrality} score than other authors ($3.8\cdot10^{-6}$ versus $2.4 \cdot 10^{-7}$).
We therefore propose that \term{eigenvector centrality} in bipartite author-paper networks is a promising approach for studying the role and impact of collaborating individuals in science, and might serve as an objective measure of influence in scientific publishing.

\subsection{Modularity-driven Clustering}

\term{Graph clustering} comprises a variety of methods for detecting natural communities in networks. Formally, it is concerned with partitioning the node set into disjoint subsets (clusters), the result of which is called a \term{clustering}.
The notion of a cluster is usually based on the \term{intra-cluster density versus inter-cluster sparsity} paradigm, according to which a clustering should identify groups of nodes which are internally densely connected, while only sparse connections exist between the groups.
One of the primary measures of clustering quality based on this paradigm is \term{modularity}~\cite{ng-fecsn-04}.

\begin{definition}
For a graph $G = (V,E)$ and a clustering $\zeta = \set{C_1, \dots, C_k}$ of $G$, \term{modularity} is defined as
\begin{align*}
\func{mod}(G, \zeta) := 
 \sum_{C \in \zeta} \frac{|E(C)|}{|E|} - \sum_{C \in \zeta}   \frac{ \left (  \sum_{v \in C} deg(v) \right )^2}{(2 \cdot |E|)^2}
\end{align*}
\end{definition}

The measure considers the clustering's \term{coverage} (the fraction of edges placed within a cluster) on the actual graph and subtracts the \term{coverage} it would achieve on a randomly connected version of the graph (preserving degree distribution).
 \term{Modularity}-based clusterings often agree with human intuition, although criticism has emerged recently~\cite{PhysRevE.84.066122}. 
Since maximizing \term{modularity} is an \NP-hard problem~\cite{bdgghnw-omc-08}, we use a heuristic based on \term{local greedy agglomeration}.
The base algorithm, commonly referred to as the \algo{Louvain Method}~\cite{bgll-f-08}, starts with a singleton clustering, considers  nodes in turn, moves them to the best neighboring cluster and contracts the graph for the next iteration.
This yields a hierarchy of graphs with increasing coarseness where the clustering in the coarsest level induces the resulting clustering in the original graph.
Rotta \etal~\cite{rn-m-11} enhance this algorithm by a refinement phase that iteratively projects this clustering to lower levels of the hierarchy and further improves modularity by local node moves. 
We use this modified algorithm.

It is a common approach to apply a clustering method to a real world network and then compare it to a ground-truth partition of the node set in order to interpret the result.
For example, Rodriguez \etal\cite{Rodriguez2008} study sensor networks research groups and apply clustering techniques like \term{leading eigenvector}, but not \term{modularity} maximization; these network-structural communities are then compared to communities defined by socio-academic similarities. 
 
 \begin{figure}[bt]
\begin{center}
\includegraphics[]{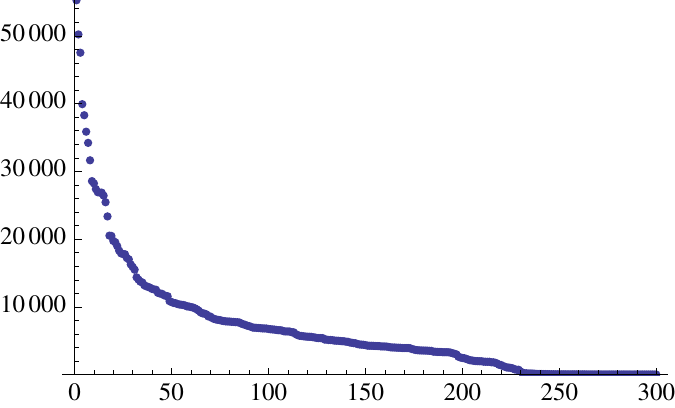}
\caption{Size distribution for the 300 largest clusters (x-axis: cluster size, y-axis: frequency)}
\label{fig:largest-clusters}
\end{center}
\end{figure}

We therefore proceed as follows:
Applying local greedy agglomeration to $G_\mb{PA}$ yields a clustering with 86761 clusters, achieving a \term{modularity} of 0.896896. 
The majority of clusters contain only a handful of nodes, and likely correspond to the many tiny components of the graph, while the dominant connected component is divided into several large clusters (see Figure~\ref{fig:largest-clusters}). 
With a clustering of the \term{authorship graph} at hand, we attempt to interpret such a \term{modularity}-driven clustering in the context of collaboration networks.
The partition found by maximizing \term{modularity} locally identifies groups of authors who are densely connected through collaborative ties.
Our hypothesis is that we can infer a topical similarity from these connections.
More precisely, we conjecture that researchers form collaborative ties around distinct areas of research, which is reflected in the clustering structure of the graph.
To put this hypothesis to the test, we compare the \term{modularity clustering}\ of $G_\mb{PA}$ to a ground-truth subdivision of the author set based on conferences: Assuming that distinct areas of computer science generally have dedicated conferences, we assign all authors who have published at a particular conference to an author-cluster.
(Unlike the \term{modularity clustering}, this does not yield a proper, complete and disjoint partition of the author set, but is nonetheless informative.)
Thereby we arrive at \term{topical clusters} of authors, which are suited as a ground-truth to compare the \term{modularity clustering} to.

\begin{table}[bp]
\begin{center}
\begin{tabular}{|c|c|c|}
\hline
& random & topical \\
\hline
$O$ & 0.04404 & 0.22832 \\
$J$ & 0.00372 & 0.01390 \\
\hline
\end{tabular}
\end{center}
\caption{Mean maximum overlap for modularity clustering and random vs topical clustering}
\label{tab:mean-max-overlap}
\end{table}%

In order to evaluate the similarity between the two community structures, one being the \term{modularity clustering}, the other the \term{topical clustering} defined by conferences, we apply overlap measures to each pair of clusters: The \term{Jaccard index} $J(A, B) := {{|A \cap B|}\over{|A \cup B|}}$ favors exact match of the two sets; the \term{overlap coefficient} $O(A, B) := {{|A \cap B|}\over{\min(|A|, |B|)}}$ treats containment of one set in the other set as a strong match, which is more equitable when dealing with clusters of uneven sizes. Applying these measures yields matrices of overlap values between \term{modularity clusters} and \term{topical clusters}. 
Additionally, we arrive at a baseline for the overlap values by calculating the overlap matrix of \term{modularity} clustering and a random clustering. The random clustering is constructed by copying the size distribution of the 250 largest modularity clusters, but randomly assigning authors to the clusters.

In these overlap matrices, we are interested in the maximum entry for each row, pointing to pairs of clusters that are most similar. Table~\ref{tab:mean-max-overlap} shows the means of these maximum overlap values. It is evident that the maximum $J$ and $O$ overlap is significantly better for \term{modularity clusters} than for random clusters. This shows that a more than coincidental relation between \term{modularity clusters} and \term{topical clusters} exists. However, the values are not close to 1.0 and indicate that the correspondence is not very strong. Thus, factors in addition to joint conferences are influential in shaping the community structure of the network. In the following section, we take an in-depth look at one possible factor of this kind, namely participation in research seminars.

\section{Impact of Seminars on Network Evolution}
\label{sec:impact}

After describing static aspects of the network in the previous section, this section is concerned with its dynamics: We examine whether the \atitle{Dagstuhl Seminars}, as academic and social events, leave a track in the structure of the network, preferably in the form of increased collaboration between the participants.
In the authors' subjective experience, the seminars present valuable opportunities for networking.
Our approach to this question can be summarized as follows: Track groups of researchers (seminar participants and others selected as reference groups) in the time-resolved graph $\mathcal{G}_{\mb{PA}}^{1,1}$ and observe their publication output as well as their collaborative links; take into account the date of a seminar in order to observe immediate or long-term effects. The preparations necessary for this approach are described in the following:

 \subsection{Preparations}
 
 \paragraph{Aligning Data Sets}

Our data sets record a total of \numprint{11625} seminar guests in the \atitle{Dagstuhl} database and \numprint{852250} authors in \atitle{DBLP}. All seminars took place in the 2000s. We align the tests by author name, whereby some false (mis)matches cannot be avoided. Still, a matching author in the publication database was found for 72 percent of the seminar invitees. 
 
 \paragraph{Area Launchers}
  
In order to detect increased collaboration which can be clearly attributed to the seminars, we first try to identify \term{area launchers}.
These are seminars intended to bring together a group of researchers who have not collaborated much before. A stated goal of the \atitle{Dagstuhl Seminars} is that some of them are intended to “launch” new areas of research by fostering collaboration between previously unaffiliated researchers, thereby contributing to emerging fields.
\term{Area launchers} are relevant to us due to the following argument: If participants develop collaborative ties in the aftermath of an \term{area launcher} seminar, it is possible to attribute this more clearly to the seminar rather than existing relationships, developed, for instance, in the course of a common conference.
 
 We classify a set of seminars as \term{area launchers} without special knowledge about the intent or content of the seminar, but solely from participation data:
  It is assumed that well-established areas of research generally spawn their own dedicated conference, and that the participants of such a conference represent the researchers active in this area. By this logic, a seminar corresponds to an established area of research if the invitees have a strong overlap with the participants of the respective conference.
Furthermore, if researchers attend the same conference, it is likely that they are already familiar with each other as well as each other's work.
We therefore reason that a seminar is an \term{area launcher} if its invitees do not overlap strongly and clearly with the participants of any particular conference. From this calculated set of seminars, 10 seminars are selected by hand and classified as \term{area launchers}.

\paragraph{Measures}

\begin{figure}[bt]
\begin{center}
\includegraphics[width=0.55\columnwidth]{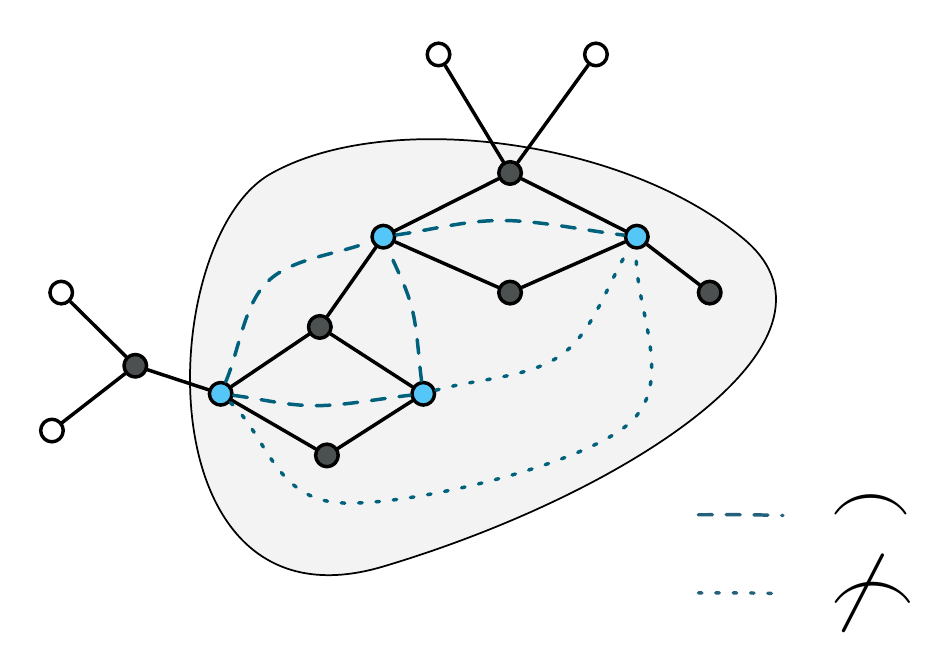}
\caption{Illustrating collaboration measure $\func{cad}$: $\func{cad}(A) = 2/3$}
\label{fig:cad}
\end{center}
\end{figure}

We quantify the publication output and intensity of collaboration among researchers using several measures which map sets of authors to real numbers. For example, Figure~\ref{fig:cad} shows a small number of authors (light nodes) and their publications (dark nodes) in the \term{authorship graph}. Authors belonging to $A$ are colored blue. Blue lines show existing (dashed line) and nonexisting (dotted line) coauthorship relations between pairs of authors in $A$. This illustrates the measure $\func{cad}(A)$, which is the fraction of actually existing coauthorship relations within an author set.
Before introducing all measures, it is helpful to define sets of (co)publications, copublications internal to a group, and coauthors first: Given a set of authors $A \subseteq \mb{A}$, the set of their publications $P(A)$ is equal to
\begin{equation*}
	P(A) := \bigcup_{a \in A} P(a) = \bigcup_{a \in A} \set{p \in \mb{P}: a \auth  p }
\end{equation*}

 The set of \term{copublications} for an author $a$ consists of publications which were written as collaborations with another author:
\begin{equation*}
	C\!P(a) := \{p \in P(a) :  \exists b \in \mb{A}: b \auth p \}
\end{equation*}
For an author set $A \subseteq \mb{A}$, the \term{aggregated copublications} are
\begin{equation*}
	C\!P(A) := \bigcup_{a \in A} C\!P(a)
\end{equation*}
%

The set of \term{intra-copublications} of a set of authors is defined as
\begin{equation*}
		C\!P_{\text{intra}}(A) := \left \{ p \in C\!P(A): \exists {a, b} \in A: a \auth p,  b \auth p \right \}
	\end{equation*}
The set of coauthors for a given author $a \in \mb{A}$ are those authors with whom $a$ has authored a collaboration.
\begin{equation*}
		C\!A(a) := \{ b \in \mb{A}: b \coauth a \} 
\end{equation*}
This can be generalized for a set of authors $A$:	
\begin{equation*}
		C\!A(A) := \bigcup_{a \in A} C\!A(a)
\end{equation*}
Based on these sets, we formulate five measures, listed and defined in Table~\ref{tab:measures}. These measures are intended to answer the following questions:
\begin{itemize}
	\item $\func{ap}(A)$: What is the general productivity of an average author from the group?
	\item $\func{acp}(A)$: What is the productivity of such an author in terms of collaborations?
	\item $\func{aca}(A)$: With how many other authors does an average author from the group collaborate?
	\item $\func{cpr}_{\text{intra}}(A)$: Do the authors collaborate more often within the group or outside of the group?
	\item $\func{cad}(A)$:  How close is the group to a collaborative clique, \ie a group in which all authors have collaborated with each other?
\end{itemize}

\setlength{\extrarowheight}{4pt}
\begin{table}[htbp]
\begin{center}
\begin{tabular}{|c|c|}
\hline
measure & definition \\
\hline
$a\!p(A)$ & $\frac{|P(A)|}{|A|}$ \\
\hline
$acp(A)$ & $\frac{|C\!P(A)|}{|A|}$ \\
\hline
$\func{aca}(A)$ & $\frac{|C\!A(A)|}{|A|}$ \\
\hline
$cpr_{\text{intra}}(A)$ & $\frac{ |  C\!P_\text{intra}(A)| }{ |C\!P(A)|}$ \\
\hline
 $cad(A)$ &  $|  \{ \{a,b\} \in \pairs{A}: a \frown b \} | / | \pairs{A} | $ \\
 \hline
\end{tabular}
\end{center}
\caption{Overview of collaboration measures and their definitions}
\label{tab:measures}
\end{table}%
\setlength{\extrarowheight}{0pt}

\paragraph{Author Classes}

The classes of author groups which we track are the seminar participants on the one hand and several reference classes on the other:

\begin{itemize}
\item \term{seminar attendees} ($A\!t_s$):  For each seminar $s$, the set of researchers who attended the seminar. 
\item \term{seminar absentees} ($A\!b_s$): For each seminar $s$, the set of researchers who were invited to the seminar but did not attend. (For some seminars, the set was empty or very small, so these are only included if they have a sufficient size.)
\item \term{random samples} ($R\!S_i$)  Contains randomly assembled sets of authors with the size of a typical seminar.
\item \term{connected samples} ($C\!S_i$) Contains sets of authors found by collecting nodes from $G_\mb{PA}$ in a breadth-first search from a random initial node until the typical size of a seminar is reached.
\item \term{all authors} ($\mb{A}$) A single set containing all authors.
\end{itemize}

\subsection{Evaluation and Results}

\setcounter{paragraph}{0}

\begin{figure}[htbp]
\begin{center}
    \subfigure[$\func{aca}$: $A\!t$]{
\includegraphics[height=29ex]{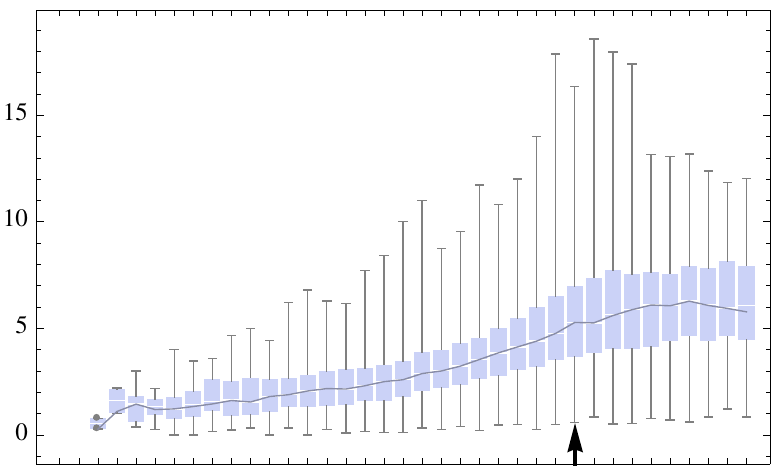}
      \label{fig:attendees}
    }
    \subfigure[$\func{aca}$: $A\!b$]{
\includegraphics[height=29ex]{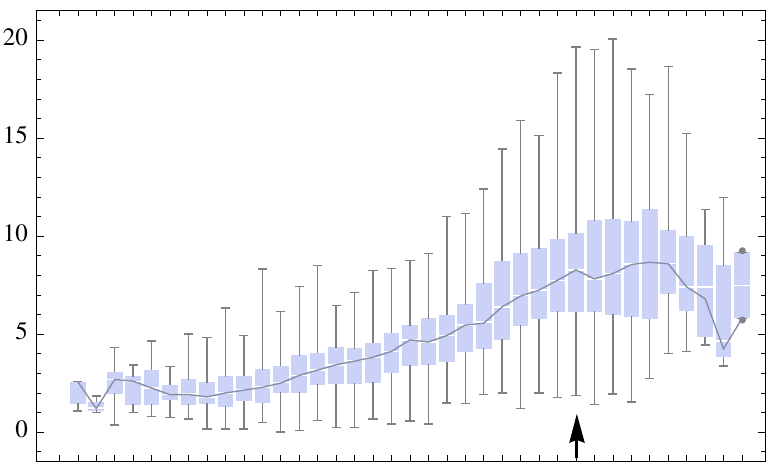}
      \label{fig:absentees-aca}
    }
\end{center}
\caption{$\func{aca}$ (y-axis) for seminar attendees and absentees (x-axis: time relative to seminar, arrow: seminar date)}
\label{fig:aca-att-ab}
\end{figure}

We speculate that joint participation in a seminar leads to increased collaboration between the participants. This would be measurable as higher values for the collaboration measures ($\func{cad}$, $\func{cpr}_{\text{intra}}$) on the respective subgraph.
Additionally, we measure whether seminar participation leads to a higher publication output for the participants ($\func{ap}$, $\func{acp}$, $\func{aca}$).
In order to test this, seminar-related groups as well as reference groups are tracked within the graph $\mathcal{G}^{1,1}_\mb{PA}$:
For any author set $A$, a subset $A' \subseteq A$ has corresponding nodes $V_{A'}$ in the graph $G_{y}$. For all measures $M$, we evaluate $M(A')$, yielding a sequence of values for each group.
The evaluation yields one value sequence per author group, and thus several data points per year.
All seminar-related sequences are aligned according to the time of the seminar, in order to compare values before and after seminar participation.
We present these data points in boxplot form (\eg Figure~\ref{fig:aca-att-ab}), with the horizontal axis denoting time relative to the seminar date and the vertical axis values of the respective measure.
By following the plotted median and quantiles along the time axis, one can identify trends for the author class as a whole.
The point in time where a seminar occurs is marked by an arrow.

In the following section, we describe a selection of notable observations:

\paragraph{Average publication output remains rather constant}

For the authors as a whole ($\mb{A}$), average publication output and number of coauthors remain stable over time, even as the graph grows at an increasing rate and author nodes accumulate. 

\paragraph{Randomly grouped authors as a baseline for publication output}

As a reference class, we evaluate the randomly compiled author groups $R\!S$. Both $\func{ap}$ and $\func{aca}$ are, on average, in the range of 0.6-0.8, showing that there are typically inactive authors in any given time frame. As expected, there is no collaboration between authors in the random samples.

\paragraph{Connected Sample Groups}

Authors from the $C\!S$ have a significantly higher productivity than randomly selected authors, since breadth-first search finds high-degree nodes with a higher probability. There is also an upward trend over time for all measures. A possible explanation for this is that nodes gain connections over time according to degree, if there is an underlying \term{preferential-attachment} process at work (as suggested by the power-law degree distribution). Overall $\func{cpr}_{\text{intra}}$ remains clearly below 0.5, showing that these sample groups are just sections from greater collaborative clusters.

\paragraph{Attendees and absentees are equally productive}

The effect of seminar participation is best judged by contrasting attendees with absentees.
With respect to productivity, measured by the number of coauthors and the number of publications, attendees and absentees are quite similar, with some outliers among the absentees surpassing the attendees (see Figure~\ref{fig:aca-att-ab}).
For the productivity measures, an upward trend before the seminar continues for a few years but then tends to reverse.

\paragraph{Attendees form a more cohesive group}

\begin{figure}[htbp]
\begin{center}
    \subfigure[$\func{cpr}_{\text{intra}}$: $A\!t$]{
\includegraphics[height=29ex]{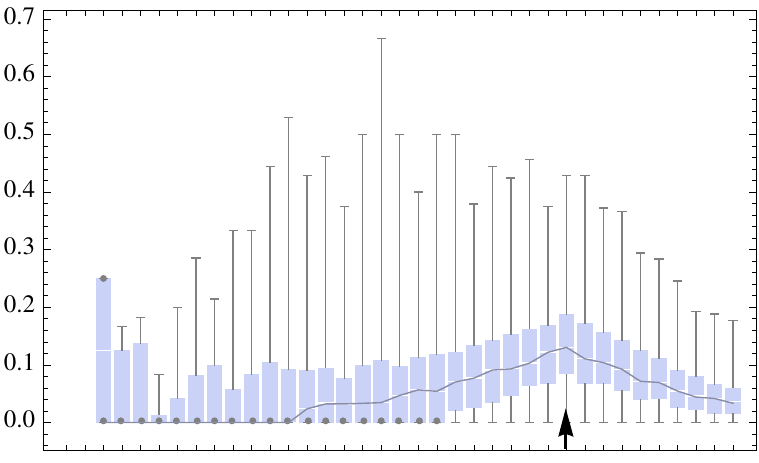}
      \label{fig:cprintra-attendees}
    }
    \subfigure[$\func{cpr}_{\text{intra}}$: $A\!b$]{
\includegraphics[height=29ex]{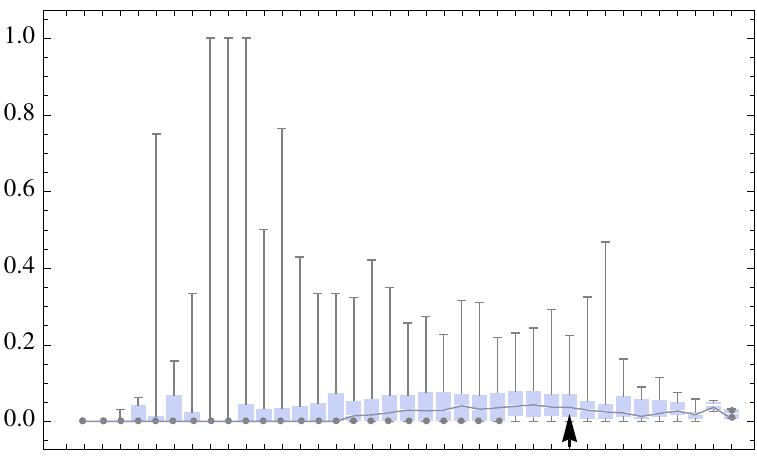}
      \label{fig:cprintra-absentees}
    }
\end{center}
\caption{$\func{cpr}_{\text{intra}}$ (y-axis) for seminar attendees and absentees (x-axis: time relative to seminar, arrow: seminar date)}
\label{fig:cprintra-att-ab}
\end{figure}

For seminar attendees, a larger fraction of their collaborations are internal to the seminar group, both before and after the seminar (Figure~\ref{fig:cprintra-att-ab}).
This indicates that attendees already come from a more cohesive group.
Values for $\func{cad}$ agree with this interpretation: Clearly, those who choose to attend the seminar form a denser subgraph in the collaboration network.
There seems to be no lasting increase in collaboration after the seminar, but a downward trend for both attendees and absentees.

\paragraph{Area launchers are not exceptional}

\begin{figure}[h]
\begin{center}
    \subfigure[$\func{cpr}_{\text{intra}}$: $A\!t$]{
\includegraphics[height=29ex]{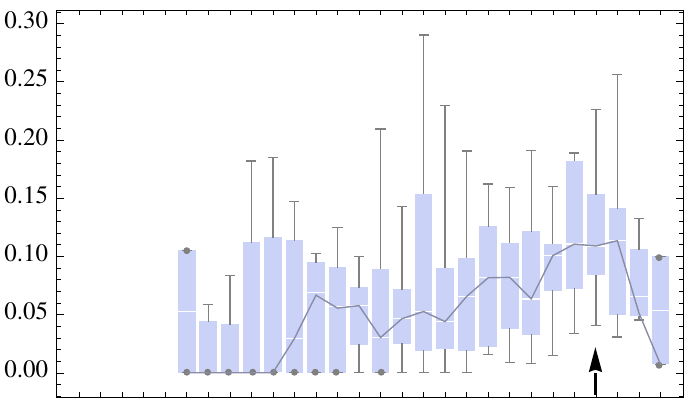}
      \label{fig:cprintra-area-attendees}
    }
    \subfigure[$\func{cpr}_{\text{intra}}$: $A\!b$]{
\includegraphics[height=29ex]{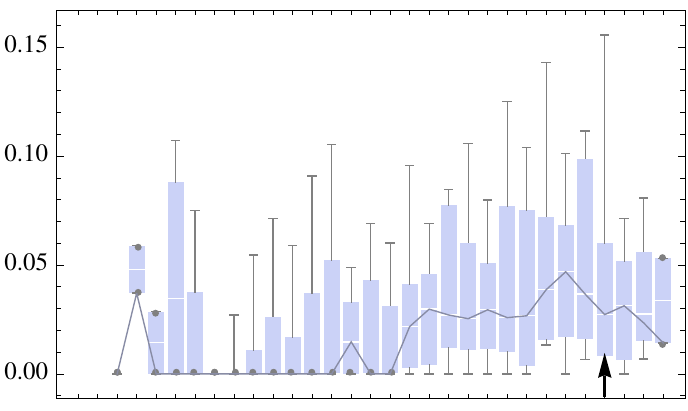}
      \label{fig:cprintra-area-absentees}
    }
\end{center}
\caption{$\func{cpr}_{\text{intra}}$ (y-axis) for attendees and absentees of \term{area launchers} (x-axis: time relative to seminar, arrow: seminar date)}
\label{fig:cprintra-area-att-ab}
\end{figure}

For the subset of seminars classified as \term{area launchers}, we expect comparatively less collaboration before the seminar, and a stronger increase after.
This effect would be most clearly captured by the measures $\func{cpr}_{\text{intra}}$ (Figure~\ref{fig:cprintra-area-att-ab}) and $\func{cad}$.
The plots in Figure~\ref{fig:cprintra-area-att-ab} support our reasoning about area launchers, namely that the authors invited have a comparatively low probability of collaboration in the time prior to the seminar:
Values for $\func{cpr}_{\text{intra}}$ are generally in the lower range compared to all seminars. Still, a visible change after the time of the seminar is missing.
The influence of an \term{area launcher} seminar does not seem to differ from the other seminars.

\paragraph{Subdivision by career stage}

Suspecting that seminar participation affects researchers in early stages of their career more strongly, we repeat a part of the evaluation with the authors classified by career length ($\le 5$, $\le 15$, $>15$ years of publication history).
However, the results do not modify our conclusions: A seminar effect for academic newcomers is no more observable than for all other authors. 

\paragraph{Summary and Interpretation}

Seminar invitees are more productive and more collaborative than randomly selected authors.
Yet there is little difference between attendees and absentees in terms of their productivity.
Invited researchers are already actively publishing, with an upward trend, prior to the time of the seminar. For $\func{cpr}_{\text{intra}}$ and $\func{cad}$, attendees are consistently better than absentees.
This indicates that those who attend are already a tightly connected collaborative group before the seminar, possibly influencing their decision to participate.
The general trend over time is an increase up to the seminar and a slight decrease afterwards for both classes of researchers.
A possible explanation for the increase and decrease over time is that invitations are biased towards researchers who are currently most active:
Invitations to seminars occur at a period of peak activity.
There is, however, no significant change of structure connected to seminars (either significant short-term increase in collaboration directly after the seminar or long-term increase).
Most importantly, attendees and absentees do not differ in this respect.
While the focus on \term{area launcher} seminars supports our assumption that the invited researchers had collaborated less, a significant structural change after the seminar is not visible.
These results suggest that a single event like a seminar is not influential enough to alter the network structure of collaboration for the group of participants in ways observable with our measures.
Clearly, other factors have additional and apparently more influence on the structure.
Rather in the opposite direction, the network structure might be employed to predict who will attend the seminar and who will decline, since the participants evidently come from a more cohesive group.

\section{Conclusion}

This paper ties in with the existing work on scientific collaboration networks and explores several new variations of network analysis methods.
The coauthorship graph in the field of computer science constitutes in many respects a typical social network, as observed before in similar studies:
We encounter properties such as low average distances between researchers, a \term{giant connected component}, a power-law distribution with regard to publications and coauthors (making it a \term{scale-free network}), and a regular \term{$k$-core} structure.
We detect dense communities of researchers through \term{modularity} maximization, and compare the resulting clustering to ground-truth communities defined by conferences, from which topical similarity is inferred. The overlap between the two partitions is clearly not coincidental, although other factors seem to be at work in shaping the community structure.
In order to identify influential researchers by their network centrality, we test a novel combination of bipartite author-paper graph and \term{eigenvector centrality}. 
We are the first to incorporate data on participants of the \atitle{Schloss Dagstuhl} research seminars and use it to evaluate the impact of such seminars on the evolution of collaborative ties.
Since the seminars are designed to foster collaboration on cutting-edge research topics, and many participants experience the seminars as a valuable opportunity for networking, we investigate whether such effects can be observed as structural changes in the collaboration network.
Seminar invitees are more productive, more collaborative and structurally prominent compared to the average researcher.
However, our methods suggest that seminar participation does not directly affect the structure of the collaboration network.
An interesting finding of this analysis was that researchers who choose to attend the seminar form a distinctly more cohesive subgraph than those who decline.


\section*{Acknowledgment}

We thank Ulrik Brandes for helpful discussions during the preparation of this work. 
We also thank the \atitle{Schloss Dagstuhl} conference center for providing us with the necessary data on their seminars.




\bibliographystyle{IEEEtran}
\bibliography{IEEEabrv,SNPaper,references,thesis}

\end{document}